\documentclass{article}
\usepackage{amssymb}
\usepackage{amsmath}
\begin{document}
\baselineskip16pt
\title{Charged Brownian particle in a magnetic field }
\author{Rados{\l}aw Czopnik \\
Institute of Theoretical Physics, University of Wroc{\l}aw, \\
PL-50 205 Wroc{\l}aw, Poland\\ and \\
 Piotr Garbaczewski\\ Institute of
Physics, Pedagogical University,\\ PL-65 069 Zielona G\'{o}ra,
Poland}

\maketitle

\begin{abstract}
We derive explicit forms  of Markovian transition probability
densities for  the velocity space and   phase-space Brownian
motion of a charged particle in a constant magnetic field.
 \end{abstract}

\section{\protect\bigskip Motivation}

An old-fashioned problem of  the Brownian motion of a charged
particle in a constant magnetic field has originated from studies
of the diffusion of plasma across a magnetic field \cite{Tay},
\cite{Kur} and nowadays, together with a free Brownian motion
example,  stands for a textbook illustration of how transport and
auto-correlation functions should be computed in generic
situations governed by the Langevin equation  cf. \cite{Bal} but
also \cite{Sch}, \cite{vKa}. To our knowledge, except for the
paper \cite{Kur} no attempt was made in the literature to give a
complete characterization of the pertinent stochastic process.
However a striking  peculiarity of  Ref. \cite{Kur} is that the
Brownian motion in a magnetic field
 is there described in terms of \it
operator-valued \rm (matrix-valued functions) probability
distributions that  involve fractional powers of matrices. In
consequence, we have  no clean  relationship with the standard
formalism of  Kramers-Smoluchowski equations, nor ways to stay in
conformity with  the standard wisdom about probabilistic
procedures valid in case of the free Brownian motion
(Ornstein-Uhlenbeck process), cf. \cite{Step}, \cite{Cha},
\cite{Nel}. Therefore, we  address an issue of the Brownian motion
of a charged particle  in a magnetic field anew, to unravel its
features of a fully-fledged stochastic diffusion process.

\section{Velocity-space diffusion process}

The standard analysis of the Brownian motion of a free particle employs
the Langevin equation
$\frac{d\overrightarrow{u}}{dt}=-\beta \overrightarrow{u}+
\overrightarrow{A}%
\left( t\right)$ where $\overrightarrow{u}$ denotes the velocity
of the particle and  the influence of the surrounding medium on
the motion (random acceleration) of the particle is modeled by
means of two independent contributions. A systematic part $-\beta
\overrightarrow{u}$ represents  a dynamical friction. The
remaining fluctuating part $\overrightarrow{A}\left( t\right) $ is
supposed to display a statistics of  the familiar white noise:
(i)$\overrightarrow{A}\left( t\right) $ is independent of $%
\overrightarrow{u}$, (ii) $\left\langle A_{i}\left( s\right)
\right\rangle = 0$ and $\left\langle A_{i}\left( s\right)
A_{j}\left( s^{\shortmid }\right) \right\rangle =2q\delta
_{ij}\delta \left( s-s^{\shortmid }\right) $ for $i,j=1,2,3$,
where $q=\frac{k_{B}T}{m}\beta $ is a physical parameter. The
Ornstein-Uhlenbeck stochastic process comes out
 on that conceptual basis.

The linear friction model  can be  adopted  to the case of
diffusion of charged particles in the presence of a constant
magnetic field which acts upon particles via the Lorentz force.
The Langevin equation for that  motion reads:

\begin{equation}
\frac{d\overrightarrow{u}}{dt}=-\beta \overrightarrow{u}+\frac{q_{e}}{mc}%
\overrightarrow{u}\times \overrightarrow{B}+\overrightarrow{A}\left( t\right)
\label{Langevin}
\end{equation}

where $q_{e}$ denotes an  electric charge of the particle of mass $m$.

Let us assume for simplicity that the constant magnetic field
$\overrightarrow{B}$ is
directed along the z-axis of a Cartesian reference frame:
$ \overrightarrow{B}=\left( 0,0,B\right) $ and $B=const$. In this case
Eq. (\ref{Langevin}) takes the form

\begin{equation}
\frac{d\overrightarrow{u}}{dt}=-\Lambda \overrightarrow{u}+
\overrightarrow{A}%
\left( t\right)  \label{LanII}
\end{equation}

where
\begin{equation}
\Lambda =\left(
\begin{array}{ccc}
\beta  & -\omega _{c} & 0 \\
\omega _{c} & \beta  & 0 \\
0 & 0 & \beta
\end{array}
\right)
\end{equation}

 and $\omega _{c}=\frac{q_{e}B}{mc}$
 denotes the Larmor frequency.
Assuming the Langevin equation to be (at least formally) solvable,
we can infer  a probability density  $P\left( \overrightarrow{u},t|%
\overrightarrow{u}_{0}\right) $, $t>0$ conditioned by the  the initial
velocity data choice $\overrightarrow{%
u}=\overrightarrow{u}_{0}$ at $t=0$. Physical circumstances of the
problem enforce a demand: (i) $P\left(
\overrightarrow{u},t|\overrightarrow{u}_{0}\right)
\rightarrow \delta ^{3}\left( \overrightarrow{u}-\overrightarrow{u}%
_{0}\right) $ as $t\rightarrow 0$ and (ii)  $P\left(
\overrightarrow{u},t|\overrightarrow{u}_{0}\right) \rightarrow
\left( \frac{m}{2\pi k_{B}T}\right) ^{\frac{3}{2}}\exp \left(
-\frac{m |\overrightarrow{u}_{0}| ^{2}}{2k_{B}T} \right) $ as
$t\rightarrow \infty $.

A  formal solution of Eq. (\ref{LanII}) reads:

\begin{equation}
\overrightarrow{u}\left( t\right) -e^{-\Lambda t}\overrightarrow{u}%
_{0}=\int_{0}^{t}e^{-\Lambda \left( t-s\right) }
\overrightarrow{A}\left( s\right) ds \enspace .  \label{sol1}
\end{equation}

By taking into account that
\begin{equation}
e^{-\Lambda t}=e^{-\beta t}\left(
\begin{array}{ccc}
\cos \omega _{c}t & \sin \omega _{c}t & 0 \\
-\sin \omega _{c}t & \cos \omega _{c}t & 0 \\
0 & 0 & 1
\end{array}
\right) =e^{-\beta t}U\left( t\right)
\end{equation}

we can rewrite (\ref{sol1}) as  follows

\begin{equation}
\overrightarrow{u}\left( t\right) -e^{-\beta t}U\left( t\right)
\overrightarrow{u}_{0}=\int_{0}^{t}e^{-\beta \left( t-s\right)
}U\left( t-s\right) \overrightarrow{A}\left( s\right) ds  \enspace
.
\end{equation}

Statistical properties of $\overrightarrow{u}\left( t\right)
-e^{-\Lambda t}\overrightarrow{u}_{0}$ are identical with those of
 $
\overrightarrow{A}\left( s\right)
ds$.
 In consequence, the problem of deducing a probability density
 $P\left(
\overrightarrow{u},t|\overrightarrow{u}_{0}\right) $ is
equivalent to deriving  the probability distribution
of the random vector

\begin{equation}
\overrightarrow{S}=\int_{0}^{t}\psi
\left( s\right) \overrightarrow{A}\left(
s\right) ds  \label{Sdef}
\end{equation}

where  $\psi \left( s\right) =e^{-\Lambda \left( t-s\right)
}=e^{-\beta \left( t-s\right) }U\left( t-s\right)$.

The white noise term $ \overrightarrow{A}\left( s\right) $ in view
of the integration with respect to time is amenable to a more
rigorous analysis that invokes the Wiener process increments and
their statistics, \cite{Doob}. Let us divide the time integration
interval into a  large number of small subintervals  $\Delta t$.
We adjust them suitably to assure  that effectively $\psi \left(
s\right) $ is constant on each subinterval $\left( j\Delta
t,\left( j+1\right) \Delta t\right) $ and equal $\psi \left(
j\Delta t\right) $. As a result we obtain the expression

\begin{equation}
\overrightarrow{S}=\sum_{j=0}^{N-1}\psi \left( j\Delta t\right)
\int_{j\Delta t}^{\left( j+1\right) \Delta
t}\overrightarrow{A}\left( s\right) ds \enspace .  \label{S}
\end{equation}

Here $ \overrightarrow{B}\left( \Delta t\right) =\int_{j\Delta
t}^{\left( j+1\right) \Delta t}\overrightarrow{A}\left( s\right)
ds$ stands for the above-mentioned Wiener process  increment.
Physically, $\overrightarrow{B}\left( \Delta t\right) $
 represents the \it net \rm acceleration which a Brownian particle
may suffer (in fact accumulates)  during an interval of time
$\Delta t$. Equation (\ref{S}) becomes

\begin{equation}
\overrightarrow{S}=\sum_{j=0}^{N-1}\psi \left( j\Delta t\right)
\overrightarrow{B}\left( \Delta t\right) =
\sum_{j=0}^{N-1}\overrightarrow{s}%
_{j}
\end{equation}

where we introduce $\overrightarrow{s}_{j}=\psi \left( j\Delta t\right)
\overrightarrow{B}\left( \Delta t\right) =
\psi _{j}\overrightarrow{B}\left(
\Delta t\right) $.

The Wiener process argument \cite{Cha}, \cite{Nel},  allows us to
infer  the probability distribution of $\overrightarrow{s}_{j}$.
It is enough  to employ the fact that the  distribution of
$\overrightarrow{B}\left( \Delta t\right) $ is Gaussian with mean
zero and variance $q=\frac{k_{B}T}{m}\beta $. Then

\begin{equation}
w\left[ \overrightarrow{B}\left( \Delta t\right) \right] =
\left( \frac{1}{%
4\pi q\Delta t}\right) ^{\frac{3}{2}}\exp \left( -\frac{\left|
\overrightarrow{B}\left(
\Delta t\right) \right| ^{2}}{4q\Delta t}\right)
\label{w(B)}
\end{equation}

and in view of $\overrightarrow{s}_{j}=\psi _{j}
\overrightarrow{B}\left( \Delta t\right) $ by performing the
change of variables in (\ref{w(B)}) we get

\begin{equation}
\widetilde{w}\left[ \overrightarrow{s}_{j}\right] =
\det \left[ \psi _{j}^{-1}%
\right] w\left[ \psi _{j}^{-1}\overrightarrow{s}_{j}\right] =
\frac{1}{\det \psi _{j}}w\left[ \psi
_{j}^{-1}\overrightarrow{s}_{j}\right] \enspace .
\end{equation}

Since $
\det \psi \left( s\right) =e^{-3\beta \left( t-s\right) } $ and
$\psi ^{-1}\left( s\right) = U\left[ -\left( t-s\right) \right]
e^{\beta \left( t-s\right) }$  we obtain

\begin{equation}
\widetilde{w}\left[ \overrightarrow{s}_{j}\right] =\left( \frac{1}{4\pi
q\Delta t}\right) ^{\frac{3}{2}}\frac{1}{e^{-3\beta \left( t-j\Delta
t\right) }}\exp \left( -\frac{\left|
e^{\beta \left( t-j\Delta t\right) }U%
\left[ -\left( t-j\Delta t\right) \right]
\overrightarrow{s}_{j}\right| ^{2}%
}{4q\Delta t}\right)
\end{equation}

and finally

\begin{equation}
\widetilde{w}\left[ \overrightarrow{s}_{j}\right] =
\left( \frac{1}{4\pi
q\Delta t}\frac{1}{e^{-2\beta \left( t-j\Delta t\right) }}
\right) ^{\frac{3}{%
2}}\exp \left( -\frac{\left| \overrightarrow{s}_{j} \right|
^{2}}{4q\Delta te^{-2\beta \left( t-j\Delta t\right) }}\right)
\enspace .
\end{equation}

Clearly, $\overrightarrow{s}_{j}$ are
mutually independent random variables whose distribution is Gaussian
with mean zero and variance $\sigma
_{j}^{2}=2q\Delta te^{-2\beta \left( t-j\Delta t\right) }$.
Hence, the probability distribution of $\overrightarrow{S}=
\sum_{j=0}^{N-1}%
\overrightarrow{s}_{j}$  is again  Gaussian with mean zero. Its
variance equals  the sum of variances of $\overrightarrow{s}_{j}$
i.e.  $ \sigma ^{2}=\sum_{j}\sigma _{j}^{2}=2q\sum_{j}\Delta
te^{-2\beta \left( t-j\Delta t\right) }$. After taking the limit
$N\rightarrow \infty $ $(\Delta t\rightarrow 0)$ we arrive at

\begin{equation}
\sigma ^{2}=2q\int_{0}^{t}dse^{-2\beta \left( t-s\right) }=
\frac{k_{B}T}{m}%
\left( 1-e^{-2\beta t}\right) \enspace .
\end{equation}

Because of  $\overrightarrow{S}=\overrightarrow{u}
\left( t\right) -e^{-\Lambda t}%
\overrightarrow{u}_{0}$ the transition probability density of the
Brownian particle velocity, conditioned by the initial data
$\overrightarrow{u}_0$  at $t_0=0$ reads

\begin{equation}
P\left( \overrightarrow{u},t|\overrightarrow{u}_{0}\right) =
\left( \frac{1}{%
2\pi \frac{k_{B}T}{m}\left( 1-e^{-2\beta t}\right) }
\right) ^{\frac{3}{2}%
}\exp \left( -\frac{\left| \overrightarrow{u}-
e^{-\Lambda t}\overrightarrow{u%
}_{0}\right| ^{2}}{2\frac{k_{B}T}{m}\left( 1-e^{-2\beta t}\right)
} \right) \enspace .
\end{equation}

The process is Markovian and time-homogeneous, hence the above
formula can be trivially extended to encompass the case of
arbitrary $t_{0}\neq 0$ : $ P\left(
\overrightarrow{u},t|\overrightarrow{u}_{0},t_{0}\right)$ arises
by substituting everywhere $t-t_0$ instead of $t$.

Physical arguments (cf. demand (ii) preceding Eq. (4)) refer to an
asymptotic probability distribution (invariant measure density)
$P(u)$  of the random variable
 $\overrightarrow{u}$  in the  Maxwell-Boltzmann form

\begin{equation}
P\left( \overrightarrow{u}\right) =\left( \frac{m}{2\pi k_{B}T}
\right) ^{\frac{3}{2}}\exp \left( -\frac{m\left|
\overrightarrow{u}\right| ^{2}}{2k_{B}T}\right) \enspace .
\end{equation}

This time-independent  probability density  together with the
time-homogeneous transition density (15) uniquely determine a
stationary Markovian stochastic process for which we can evaluate
various mean values. Expectation values of velocity components
vanish: $ \left\langle u_{i}\left( t\right) \right\rangle =
\int_{-\infty }^{\infty }u_{i}P\left( \overrightarrow{u}\right)
d\overrightarrow{u}=0 $ for $i=1,2,3$. The matrix of the second
moments (velocity auto-correlation functions) reads

\begin{equation}
 \left\langle
u_{i}\left( t\right) u_{j}\left( t_{0}\right) \right\rangle
=\int_{-\infty }^{\infty }u_{i}u_{j}^{0}P\left( \overrightarrow{u},t;%
\overrightarrow{u}_{0},t_{0}\right)
d\overrightarrow{u}d\overrightarrow{u}%
_{0}
\end{equation}

 where $i,j=1,2,3$ and in view of $ P\left(
\overrightarrow{u},t;\overrightarrow{u}_{0},t_{0}\right) = P\left(
\overrightarrow{u},t|\overrightarrow{u}_{0},t_{0}\right) P\left(
\overrightarrow{u}_{0}\right)$ we arrive at the compact expression

\begin{equation}
\frac{k_{B}T}{m}e^{-\Lambda \left| t-t_{0}\right| }=\frac{k_{B}T}{m}%
e^{-\beta \left| t-t_{0}\right| }\left(
\begin{array}{ccc}
\cos \omega _{c}\left| t-t_{0}\right| & \sin
\omega _{c}\left| t-t_{0}\right|
& 0 \\
-\sin \omega _{c}\left| t-t_{0}\right| & \cos \omega _{c}\left|
t-t_{0}\right| & 0 \\
0 & 0 & 1
\end{array}
\right) \enspace .
\end{equation}

In particular, the auto-correlation function (second moment) of
the $x$-component of  velocity equals
\begin{equation}
\left\langle u_{1}\left( t\right) u_{1}\left( t_{0}\right) \right\rangle =%
\frac{k_{B}T}{m}e^{-\beta \left| t-t_{0}\right| }\cos \omega _{c}\left|
t-t_{0}\right|   \label{autocor}
\end{equation}

in  agreement with white noise calculations of Refs. \cite{Tay}
and \cite{Bal}, cf. Chap.11, formula  (11.25). The so-called
running diffusion coefficient arises here via straightforward
integration of the function $R_{11}(\tau )= <u_1(t)u_1(t_0)>$
where $\tau = t-t_0 >0$:

\begin{equation}
D_1(t) = \int_0^t <u_1(0)u_1(\tau)> d\tau  = %
{{k_BT}\over m} {{\beta + [\omega _csin(\omega _ct) - \beta
cos(\omega _ct)]exp(-\beta t)}\over {\beta ^2 +  {\omega _c}^2}}
\end{equation}
with an obvious asymptotics (the same for $D_2(t)$):
$D_B=lim_{t\rightarrow \infty } D_1(t)= {{k_BT}\over m} {\beta
\over {\beta ^2 + {\omega _c}^2}}$ and the large friction ($\omega
_c$ fixed and bounded) version $D= {{k_BT}\over {m\beta }}$.

\section{Spatial process - dynamics in the plane}

The cylindrical symmetry of the problem allows us to consider
separately processes running on  the $XY$  plane  and along the
$Z$-axis (where the free Brownian motion takes place). We shall
confine further attention to the two-dimensional $XY$-plane
problem.
 Henceforth,  each vector will carry two components which
 correspond to the $x$ and $y$ coordinates respectively.
 We will directly refer to the
vector and matrix quantities introduced in the previous section,
but while keeping the same notation, we shall  simply disregard
their $z$-coordinate contributions.

We  define the spatial  displacement $\overrightarrow{r}$ of the
Brownian particle as folows

\begin{equation}
\overrightarrow{r}-\overrightarrow{r}_{0}=\int_{0}^{t}
\overrightarrow{u}%
\left( \eta \right) dn
\end{equation}

where $\overrightarrow{u}\left( t\right) $ is given by Eq.
(\ref{LanII}) (except for disregarding the third coordinate).

Our aim is to derive the probability distribution of
$\overrightarrow{r}$\ at time $t$ provided  that the particle
position and velocity were equal
  $\overrightarrow{r}_{0}$\ and
$\overrightarrow{u}_{0}$\ respectively, at time $t_{0}=0$. To that
end we shall mimic  procedures of the previous section. In view
of:

\begin{equation}
\overrightarrow{r}-\overrightarrow{r}_{0}-
\int_{0}^{t}e^{-\Lambda \eta }%
\overrightarrow{u}_{0}=\int_{0}^{t}d\eta
\int_{0}^{\eta }dse^{-\Lambda
\left( \eta -s\right) }\overrightarrow{A}\left( s\right)
\end{equation}

we have

\begin{equation}
\overrightarrow{r}-\overrightarrow{r}_{0}-
\Lambda ^{-1}\left( 1-e^{-\Lambda
t}\right) \overrightarrow{u}_{0}=\int_{0}^{t}\Lambda ^{-1}\left(
1-e^{\Lambda \left( s-t\right) }\right)
\overrightarrow{A}\left( s\right) ds
\end{equation}

where
\begin{equation}
\Lambda ^{-1}=\frac{1}{\beta ^{2}+\omega _{c}^{2}}\left(
\begin{array}{cc}
\beta & \omega _{c} \\
-\omega _{c} & \beta
\end{array}
\right)
\end{equation}

is the inverse  of  the matrix $\Lambda $ (regarded as a
rank two  sub-matrix   of that originally introduced in
 Eq. (3)).
Let us define  two auxiliary matrices

\begin{eqnarray}
\Omega &\equiv &\Lambda ^{-1}\left( 1-e^{-\Lambda t}\right)
 \label{omega} \\
\phi \left( s\right) &\equiv &\Lambda ^{-1}\left( 1-e^{\Lambda
\left( s-t\right) }\right)  \notag \enspace .
\end{eqnarray}

Because of:

\begin{equation}
e^{-\Lambda t}=\exp \left\{ - t \left(
\begin{array}{cc}
\beta & -\omega _{c} \\
\omega _{c} & \beta
\end{array}
\right) \right\} =e^{-\beta t}\left(
\begin{array}{cc}
\cos \omega _{c}t & \sin \omega _{c}t \\
-\sin \omega _{c}t & \cos \omega _{c}t
\end{array}
\right) =e^{-\beta t}U\left( t\right)
\end{equation}

we can represent  matrices  $\Omega $,  $\phi \left( s\right) $ in
more detailed   form. We have:

\begin{equation}
\Omega =\frac{1}{\beta ^{2}+\omega
_{c}^{2}}\left\{ \left(
\begin{array}{cc}
\beta & \omega _{c} \\
-\omega _{c} & \beta
\end{array}
\right) -e^{-\beta t}\left(
\begin{array}{cc}
\beta & \omega _{c} \\
-\omega _{c} & \beta
\end{array}
\right) \left(
\begin{array}{cc}
\cos \omega _{c}t & \sin \omega _{c}t \\
-\sin \omega _{c}t & \cos \omega _{c}t
\end{array}
\right) \right\}
\end{equation}

and

\begin{equation}
\phi \left( s\right) =\Lambda ^{-1}\left( 1-e^{-\beta \left(
t-s\right) }U\left( t-s\right) \right) =
\end{equation}
\begin{equation*}
\frac{1}{\beta ^{2}+\omega _{c}^{2}}\left(
\begin{array}{cc}
\beta & \omega _{c} \\
-\omega _{c} & \beta
\end{array}
\right) \left(
\begin{array}{cc}
1-e^{\beta \left( s-t\right) }\cos \omega _{c}\left( s-t\right) &
-e^{\beta
\left( s-t\right) }\sin \omega _{c}\left( s-t\right) \\
e^{\beta \left( s-t\right) }\sin \omega _{c}\left( s-t\right) &
1-e^{\beta
\left( s-t\right) }\cos \omega _{c}\left( s-t\right)
\end{array}
\right) \enspace .
\end{equation*}

Next steps imitate procedures of the previous section. Thus, we
seek for the probability distribution of the random (planar)
vector $ \overrightarrow{R}=\int_{0}^{t}\phi \left( s\right)
\overrightarrow{A}\left( s\right) ds  \label{Rdef}$ where
$\overrightarrow{R}=\overrightarrow{r}-\overrightarrow{r}_{0}-\Omega
\overrightarrow{u}_{0}$.

Dividing the time interval $\left( 0,t\right) $\ into  small
subintervals  to  assure that $\phi \left( s\right) $ can be
regarded  constant over the time span  $\left( j\Delta t,\left(
j+1\right) \Delta t\right) $\ and equal  $\phi \left( j\Delta
t\right) $,  we obtain

\begin{equation}
\overrightarrow{R}=\sum_{j=0}^{N-1}\phi \left( j\Delta t\right)
\int_{j\Delta t}^{\left( j+1\right) \Delta
t}\overrightarrow{A}\left(
s\right) ds=\sum_{j=0}^{N-1}\phi \left( j\Delta t\right)
\overrightarrow{B}%
\left( \Delta t\right) =\sum_{j=0}^{N-1}\overrightarrow{r}_{j}
\end{equation}

where  $\overrightarrow{r}_{j}=\phi \left( j\Delta t\right)
\overrightarrow{B}\left( \Delta t\right) =\phi
_{j}\overrightarrow{B}\left( \Delta t\right) $.

By invoking the probability distribution (10) we  perform an
appropriate change of variables: $\overrightarrow{r%
}_{j}=\phi _{j}\overrightarrow{B}\left( \Delta t\right) $ to yield
a probability  distribution of  $\overrightarrow{r}_{j}$

\begin{equation}
\widetilde{w}\left[ \overrightarrow{r}_{j}\right] =
\det \left[ \phi _{j}^{-1}%
\right] w\left[ \phi _{j}^{-1}\overrightarrow{r}_{j}\right]
=\frac{1}{\det \phi _{j}}w\left[ \phi
_{j}^{-1}\overrightarrow{r}_{j}\right] \enspace .
\end{equation}

Presently (not to be confused with previous steps (11)-(15)) we
have

\begin{equation}
\det \phi \left( s\right) =\frac{1}{\beta ^{2}+\omega
_{c}^{2}}\left( 1+e^{2\beta \left( s-t\right) }-2e^{\beta \left(
s-t\right) }\cos \omega _{c}\left( s-t\right) \right)
\end{equation}

and

\begin{equation}
\phi ^{-1}\left( s\right) =\frac{1}{1+e^{2\beta \left( s-t\right)
}-2e^{\beta \left( s-t\right) }\cos \omega _{c}\left( s-t\right)
}\left[ 1-e^{\beta \left( s-t\right) }U\left( -\left( s-t\right)
\right) \right] \Lambda  \enspace .
\end{equation}

So,  the inverse of the matrix  $\phi _{j}$ has the form:

\begin{equation}
\phi _{j}^{-1}=\frac{\widetilde{A}_{j}}{\gamma _{j}}
\end{equation}

where

\begin{equation}
\widetilde{A}_{j}=\left(
\begin{array}{cc}
1-e^{\beta \left( j\Delta t-t\right) }\cos \omega _{c}\left( j\Delta
t-t\right) & e^{\beta \left( j\Delta t-t\right) }\sin \omega _{c}\left(
j\Delta t-t\right) \\
-e^{\beta \left( j\Delta t-t\right) }\sin \omega _{c}\left( j\Delta
t-t\right) & 1-e^{\beta \left( j\Delta t-t\right) }\cos
\omega _{c}\left(
j\Delta t-t\right)
\end{array}
\right) \left(
\begin{array}{cc}
\beta & -\omega _{c} \\
\omega _{c} & \beta
\end{array}
\right)
\end{equation}

and

\begin{equation}
\gamma _{j}=1+e^{2\beta \left( j\Delta t-t\right) }-2e^{\beta
\left( j\Delta t-t\right) }\cos \omega _{c}\left( j\Delta
t-t\right) \enspace .
\end{equation}

There holds:

\begin{equation}
\det \phi _{j}^{-1}=\left( \det \phi _{j}\right) ^{-1}=\left(
\beta ^{2}+\omega _{c}^{2}\right) \frac{1}{\gamma _{j}}
\end{equation}

and as a consequence  we arrive at the following  probability
distribution of $\overrightarrow{r}_{j}$

\begin{equation}
\widetilde{w}\left[ \overrightarrow{r}_{j}\right] =
\frac{1}{\frac{1}{\beta
^{2}+\omega _{c}^{2}}\gamma _{j}}\left( \frac{1}{4\pi q\Delta t}
\right) \exp
\left\{ \frac{\left| \widetilde{A}_{j}\left(
\begin{array}{c}
r_{j}^{x} \\
r_{j}^{y}
\end{array}
\right) \right| ^{2}}{\gamma _{j}^{2}4q\Delta t}\right\} \enspace
.
\end{equation}

In view of

\begin{equation}
\left| \widetilde{A}_{j}\left(
\begin{array}{c}
r_{j}^{x} \\
r_{j}^{x}
\end{array}
\right) \right| ^{2}=\left( \beta ^{2}+
\omega _{c}^{2}\right) \gamma _{j}%
\left[ \left( r_{j}^{x}\right) ^{2}+
\left( r_{j}^{y}\right) ^{2}\right]
\end{equation}

 that finally leads to

\begin{equation}
\widetilde{w}\left[ \overrightarrow{r}_{j}\right] = \left(
\frac{\beta ^2 + \omega _{c}^2}{4\pi q\Delta t \gamma _{j}}\right)
\exp \left\{ -\frac{(\beta ^2 + \omega _{c}^2)\,  \left|
\overrightarrow{r}_{j}\right| ^{2}} {4q\Delta t \gamma
_{j}}\right\} \enspace .
\end{equation}

Since this probability distribution is Gaussian with mean zero
and variance $%
\sigma _{j}^{2}=$ $2q\Delta t\frac{1}{\beta ^{2}+\omega _{c}^{2}}
\gamma _{j}$, the random vector$\ \overrightarrow{R}$ as a sum of
independent random variables $\overrightarrow{r}_{j}$ has the
distribution

\begin{equation}
w\left( \overrightarrow{R}\right) =\frac{1}{2\pi \sum_{j}
\sigma _{j}^{2}}%
\exp \left( -\frac{R_{x}^{2}+R_{y}^{2}}{2\sum_{j} \sigma
_{j}^{2}}\right) \enspace .
\end{equation}

\begin{equation}
\sigma ^{2}=\sum_{j}\sigma _{j}^{2}=2q\sum_{j}\Delta
t\frac{1}{\beta ^{2}+\omega _{c}^{2}}\gamma _{j} \enspace .
\end{equation}

In the limit of $\Delta t\rightarrow 0$ we arrive at the
integral

\begin{equation}
\sigma ^{2}=2q\frac{1}{\beta ^{2}+\omega _{c}^{2}}
\int_{0}^{t}\gamma \left(
s\right) ds
\end{equation}

with $ \int_{0}^{t}\gamma \left( s\right) ds=t+  \Theta $, where

\begin{equation}
\Theta = \Theta (t) = \frac{1}{2\beta }\left( 1-e^{-2\beta
t}\right) -2\frac{1}{\beta ^{2}+\omega _{c}^{2}}\left[ \beta
+\left( \omega _{c}\sin \omega _{c}t-\beta \cos \omega
_{c}t\right) e^{-\beta t}\right]      \enspace .
\end{equation}

That gives rise to  an ultimate form of the transition probability
density of the spatial  displacement process:

\begin{equation}
P\left( \overrightarrow{r},t|\overrightarrow{r}_{0},t_{0}=0,
\overrightarrow{u%
}_{0}\right) =\frac{1}{4\pi \frac{k_{B}T}{m}\frac{\beta }
{\beta ^{2}+\omega
_{c}^{2}}\left( t+\Theta \right) }\exp \left( -
\frac{\left| \overrightarrow{r%
}-\overrightarrow{r}_{0}-\Omega \overrightarrow{u}_{0}
\right| ^{2}}{4\frac{%
k_{B}T}{m}\frac{\beta }{\beta ^{2}+\omega _{c}^{2}}
\left( t+\Theta \right) }%
\right)
\end{equation}

with  $\Omega =\Omega (t)$  defined in Eqs. (\ref{omega}), (27).
Notice that an asymptotic diffusion coefficient
$D_B=D{\beta^2\over {\beta ^2 + \omega ^2_c}}$ encodes an
attenuation signature for the spatial dispersion (when $\omega _c$
grows up at $\beta $ fixed).

The spatial displacement process governed by the above transition
probability density surely is \it not \rm Markovian. That can be
checked by inspection:  the Chapman-Kolmogorov identity is not
valid, like in  the standard free Brownian motion example where
the Ornstein-Uhlenbeck process induced (sole) spatial dynamics is
non-Markovian as well.

\section{Phase-space process}

\subsection{Axial direction}

We take advantage of  the cylindrical symmetry of our problem, and
consider separately the (free) Brownian  dynamics in the direction
parallel to the magnetic field vector, e.g. along the $Z$-axis.

That amounts to a  familiar Ornstein-Uhlenbeck process  in its
extended phase-space form. In the absence of external forces, the
kinetic (Kramers-Fokker-Planck equation) reads:

\begin{equation}
 {\partial _t W + u\nabla _zW = \beta \nabla _u(Wu) + q
\triangle _uW} \end{equation}

where $q=D\beta ^2$.   Here  $\beta $ is the friction coefficient,
$D$ will be identified later with the spatial diffusion constant,
and (as before) we set $D=k_BT/m\beta $ in conformity with  the
Einstein fluctuation-dissipation identity. The  joint probability
distribution (in fact, density) $W(z,u,t)$ for a freely moving
Brownian particle which at $t=0$ initiates its motion at $x_0$
with an arbitrary inital velocity $u_0$ can  be given in the form
of the maximally symmetric displacement probability law:

\begin{equation}
W(z,u,t)= W(R,S) = [4\pi ^2(FG-H^2)]^{-1/2} \cdot  exp\{ - {{GR^2
- HRS + FS^2}\over {2(FG - H^2)}}\}
\end{equation}

where $R=z-u_0(1-e^{-\beta t})\beta ^{-1}$, $S=u-u_0e^{-\beta t}$
while $ F = {D\over \beta }(2\beta t - 3 +4e^{-\beta t}-
e^{-2\beta t})$\, $G=D\beta (1-e^{-2\beta t})$ and
$H=D(1-e^{-\beta t})^2$.

\subsection{Planar process}

Now we shall consider  Brownian dynamics in the direction
perpendicular to the magnetic  field $\overrightarrow{B}$, hence
(while in terms of configuration-space variables) we address an
issue of  the planar dynamics. We are  interested in the complete
phase-space process, hence we need  to specify the transition
probability density \ $P\left(
\overrightarrow{r},\overrightarrow{u},t|\overrightarrow{r}_{0},%
\overrightarrow{u}_{0},t_{0}=0\right) $ of the Markov
 process conditioned by the initial  data
 $\overrightarrow{u}=%
\overrightarrow{u}_{0}$ and $\overrightarrow{r}=
\overrightarrow{r}_{0}$ at time $%
t_{0}=0$. That  is equivalent to deducing the joint probability
distribution
$W\left( \overrightarrow{S,}%
\overrightarrow{R}\right) $\ of random vectors
$\overrightarrow{S}$ and $%
\overrightarrow{R}$, previously defined
to appear in the form $\overrightarrow{S}=\overrightarrow{u}
\left( t\right) -e^{-\Lambda t}%
\overrightarrow{u}_{0}$ and
$\overrightarrow{R}=\overrightarrow{r}-\overrightarrow{r}_{0}-
\Omega \overrightarrow{u}_{0}$   respectively. Let us stress that
presently, all vectors are regarded as two-dimensional versions
(the third component being simply disregarded) of the original
random variables we have discussed so far in Sections 2 and 3.

 Vectors  $\overrightarrow{S}$ and
$\overrightarrow{R}$ both share
 a Gaussian distribution with mean zero. Consequently,
  the joint distribution
$W\left( \overrightarrow{S,}\overrightarrow{R}\right) $\ is
determined by the matrix of variances and covariances: $C = \left(
c_{ij}\right) = \left(\left\langle x_{i}x_{j}\right\rangle
\right)$, where  we abbreviate four phase-space variables in a
single notion of  $x=\left( S_{1},S_{2},R_{1},R_{2}\right) $ and
label components of $x$ by $i,j=1,2,3,4$.  In terms of
$\overrightarrow{R}$ and $\overrightarrow{S}$ the covariance
matrix $C$ reads:

\begin{equation}
C=\left(
\begin{array}{cccc}
\left\langle S_{1}S_{1}\right\rangle &
\left\langle S_{1}S_{2}\right\rangle
& \left\langle S_{1}R_{1}\right\rangle &
\left\langle S_{1}R_{2}\right\rangle
\\
\left\langle S_{2}S_{1}\right\rangle &
\left\langle S_{2}S_{2}\right\rangle
& \left\langle S_{2}R_{1}\right\rangle &
\left\langle S_{2}R_{2}\right\rangle
\\
\left\langle R_{1}S_{1}\right\rangle &
\left\langle R_{1}S_{2}\right\rangle
& \left\langle R_{1}R_{1}\right\rangle &
\left\langle R_{1}R_{2}\right\rangle
\\
\left\langle R_{2}S_{1}\right\rangle &
\left\langle R_{2}S_{2}\right\rangle
& \left\langle R_{2}R_{1}\right\rangle &
\left\langle R_{2}R_{2}\right\rangle
\end{array}
\right) \enspace .
\end{equation}

The joint probability distribution of $\overrightarrow{S}$ and $%
\overrightarrow{R}$ is  given by

\begin{equation}
W\left( \overrightarrow{S,}\overrightarrow{R}\right) =W\left(
\overrightarrow{x}\right) =\frac{1}{4\pi ^{2}} \left(
\frac{1}{\det C}\right)^{\frac{1}{2}}\exp
\left(
-\frac{1}{2}\sum_{i,j}c_{ij}^{-1}x_{i}x_{j}\right)
\end{equation}

where $c_{ij}^{-1}$denotes the component of the inverse matrix
$C^{-1}$.

The probability distributions of $\overrightarrow{S}$ and
$\overrightarrow{R} $, which were established in the previous
sections, determine a number of expectation values:
\begin{equation}
g\equiv \left\langle S_{1}S_{1}\right\rangle =\left\langle
S_{2}S_{2}\right\rangle =\frac{k_{B}T}{m}\left( 1-e^{-2\beta t}\right)
\end{equation}

 while $ \left\langle S_{1}S_{2}\right\rangle = \left\langle
S_{2}S_{1}\right\rangle =0$. Furthermore:

\begin{equation}
f\equiv \left\langle R_{1}R_{1}\right\rangle =\left\langle
R_{2}R_{2}\right\rangle =2\frac{k_{B}T}{m}\frac{\beta } {\beta
^{2}+\omega _{c}^{2}}\left( t+\Theta \right)= 2D_B (t+\Theta ) \,
.
\end{equation}

 In addition we have
 $ \left\langle R_{1}R_{2}\right\rangle
=\left\langle R_{2}R_{1}\right\rangle =0$. As a  consequence, we
are left with only   four non-vanishing components of the
covariance matrix $C$: $ c_{13}=c_{31}=\left\langle
S_{1}R_{1}\right\rangle $, $ c_{14}=c_{41}=\left\langle
S_{1}R_{2}\right\rangle $, $ c_{23}=c_{32}=\left\langle
S_{2}R_{1}\right\rangle $,  $ c_{24}=c_{42}=\left\langle
S_{2}R_{2}\right\rangle $ which need a closer examination.

We can obtain those covariances by exploiting
a dependence of  the random quantities $%
\overrightarrow{S}$ and $\overrightarrow{R}$  on the white-noise
term $%
\overrightarrow{A}\left( s\right) $ whose statistical properties
are known. There follows:

\begin{equation}
S_{1}=\int_{0}^{t}dse^{-\beta \left( t-s\right) }\left[ \cos \omega
_{c}\left( t-s\right) A_{1}\left( s\right) +\sin \omega _{c}\left(
t-s\right) A_{2}\left( s\right) \right]
\end{equation}

\begin{equation*}
S_{2}=\int_{0}^{t}dse^{-\beta \left( t-s\right) }\left[ -\sin \omega
_{c}\left( t-s\right) A_{1}\left( s\right) +\cos \omega _{c}\left(
t-s\right) A_{2}\left( s\right) \right]
\end{equation*}

\begin{eqnarray*}
R_{1} &=&\int_{0}^{t}ds\frac{1}{\beta ^{2}+\omega _{c}^{2}}\left[
\beta \left( 1-e^{-\beta \left( t-s\right) }\cos \omega _{c}\left(
t-s\right) \right) +\omega _{c}e^{-\beta \left( t-s\right) }\sin
\omega _{c}\left( t-s\right) \right] A_{1}\left( s\right) + \\ &
&\int_{0}^{t}ds\frac{1}{\beta ^{2}+\omega _{c}^{2}}\left[ -\beta
e^{-\beta \left( t-s\right) }\sin \omega _{c}\left( t-s\right)
+\omega _{c}\left( 1-e^{-\beta \left( t-s\right) }\cos \omega _{c}
\left( t-s\right) \right) %
\right] A_{2}\left( s\right)
\end{eqnarray*}

\begin{eqnarray*}
R_{2} &=&\int_{0}^{t}ds\frac{1}{\beta ^{2}+\omega _{c}^{2}} \left[
-\omega _{c}\left( 1-e^{-\beta \left( t-s\right) } \cos \omega
_{c}\left( t-s\right) \right) +\beta e^{-\beta \left( t-s\right)
}\sin \omega _{c}\left( t-s\right) \right] A_{1}\left( s\right) +
\\ &&\int_{0}^{t}ds\frac{1}{\beta ^{2}+\omega _{c}^{2}}\left[
\omega _{c}e^{-\beta \left( t-s\right) }\sin \omega _{c} \left(
t-s\right) +\beta \left( 1-e^{-\beta \left( t-s\right) }\cos
\omega _{c} \left( t-s\right) \right) \right] A_{2}\left( s\right)
\enspace .
\end{eqnarray*}

Multiplying together suitable  components of  vectors
$\overrightarrow{S}$ and $\overrightarrow{R}$  and taking averages
of those products in conformity with
the rules $\left\langle A_{i}\left( s\right) \right\rangle =0$ and $%
\left\langle A_{i}\left( s\right) A_{j}\left( s^{\shortmid }\right)
\right\rangle =2q\delta _{ij}\delta \left( s-s^{\shortmid }\right) $,
where $%
q=\frac{k_{B}T}{m}\beta $, $i,j=1,2,3$, we arrive at:

\begin{equation}
h\equiv \left\langle R_{1}S_{1}\right\rangle = \left\langle
R_{2}S_{2}\right\rangle =2q\frac{1}{\beta ^{2}+ \omega _{c}^{2}}
\int_{0}^{t}ds e^{-\beta \left( t-s\right) } [ \beta \cos \omega
_{c}\left( t-s\right) +
\end{equation}
\begin{equation*}
\omega _{c}\sin \omega _{c} \left( t-s\right) -\beta e^{-\beta
\left( t-s\right) }]  =q\frac{1}{\beta ^{2}+\omega _{c}^{2}}
\left( 1-2e^{-\beta t}\cos \omega _{c}t+e^{-2\beta t}\right)
\end{equation*}

and
\begin{equation}
k\equiv \left\langle R_{1}S_{2}\right\rangle =-\left\langle
R_{2}S_{1}\right\rangle =2q\frac{1}{\beta ^{2}+\omega _{c}^{2}}%
\int_{0}^{t}dse^{-\beta \left( t-s\right) }[ -\beta \sin \omega
_{c}\left( t-s\right) +
\end{equation}
\begin{equation*}
\omega _{c}\cos \omega _{c} \left( t-s\right) -\omega
_{c}e^{-\beta \left( t-s\right) }] = q\frac{1}{\beta ^{2}+\omega
_{c}^{2}}\left[ 2e^{-\beta t}\sin \omega _{c}t-\frac{\omega
_{c}}{\beta }\left( 1-e^{-2\beta t}\right) \right] \enspace .
\end{equation*}

The covariance matrix $C=\left( c_{ij}\right) $ has thus the form

\begin{equation}
C=\left(
\begin{array}{cccc}
g & 0 & h & -k \\
0 & g & k & h \\
h & k & f & 0 \\
-k & h & 0 & f
\end{array}
\right)
\end{equation}

while its  inverse  $C^{-1}$ reads as follows:

\begin{equation}
C^{-1}=\frac{1}{\det C}\left( fg-h^{2}-k^{2}\right) \left(
\begin{array}{cccc}
f & 0 & -h & k \\
0 & f & -k & -h \\
-h & -k & g & 0 \\
k & -h & 0 & g
\end{array}
\right)
\end{equation}

where $ \det C=\left( fg-h^{2}-k^{2}\right) ^{2}$.
The joint probability distribution of $\overrightarrow{S}$ and $%
\overrightarrow{R}$\ can be ultimately written in the  form:

\begin{equation}
W\left( \overrightarrow{S},\overrightarrow{R}\right) =
\end{equation}
\begin{equation}
\frac{1}{4\pi ^{2}\left( fg-h^{2}-k^{2}\right) }\exp \left( -
\frac{f\left| \overrightarrow{%
S}\right| ^{2}+g\left| \overrightarrow{R}\right| ^{2}-
2h\overrightarrow{S}%
\cdot \overrightarrow{R}+2k\left( \overrightarrow{S}
\times \overrightarrow{R}%
\right) _{i=3}}{2\left( fg-h^{2}-k^{2}\right) }\right) \enspace .
\end{equation}

In the above, all vector entries are two-dimensional. The specific
 $i=3$ vector product coordinate in the exponent is simply an
abbreviation for the (ordinary $R^3$-vector product) procedure
that involves merely  first two components of three-dimensional
vectors (the third is then arbitrary and irrelevant), hence
effectively involves  our two-dimensional $\overrightarrow{R}$ and
$\overrightarrow{S}$.\\

{\bf Acknowledgement:} One of the authors (P. G.) receives
financial support from the KBN research grant No. 2 PO3B 086 16.


\begin{thebibliography}{99}


\bibitem{Tay}  J. B. Taylor, Phys. Rev. Lett. \textbf{6}, 262, (1961)

\bibitem{Kur} B. Kur\c{s}uno\v{g}lu, Ann. Phys. \textbf{17}, 259, (1962)

\bibitem{Bal}  R. Balescu, \textit{Statistical Dynamics. Matter out of
Equilibrium}. (Imperial College Press, London, 1997)

\bibitem{Sch}  Z. Schuss, \textit{Theory and Applications of Stochastic
Differential Equations}, (Wiley, NY, 1980)

\bibitem{vKa}  N. G. van Kampen, \textit{Stochastic Processes in
Physics and Chemistry}, (North Holland, Amsterdam, 1981)



\bibitem{Step} S. Stepanov, Phys. Rev. E \textbf{54}, 2209, (1996)




\bibitem{Cha} S. Chandrasekhar, Rev. Mod. Phys. \textbf{15}, 1, (1943)


\bibitem{Nel}  E. Nelson, \textit{Dynamical Theories of Brownian Motion},
(Princeton University Press, Princeton, 1967)

\bibitem{Doob} J. L. Doob, Ann. Math. \textbf{43}, 351, (1942)
\end{thebibliography}
\end{document}